\documentclass[12pt]{article}

\usepackage{graphicx}
\usepackage{bm}
\usepackage{hyperref}

\newcommand{\eq}[1]{Eq.~(\ref{#1})}

\newcommand{\dpar}[2]{\frac{\partial#1}{\partial#2}}
\def\be{\begin{equation}}
\def\ee{\end{equation}}
\def\ba{\begin{eqnarray}}
\def\ea{\end{eqnarray}}
\def\half{\frac{1}{2}}
\def\part{\partial}

\def\a{\alpha}

\def\lam{\lambda}

\def\eps{\epsilon}

\def\bmp{\bm{p}}
\def\bmpi{\bm{\pi}}

\def\bmb{\bm{b}}
\def\bmg{\bm{g}}
\def\n{\bm{n}}
\def\bpsi{\bm{\psi}}
\def\tbpsi{\tilde{\bm{\psi}}}
\def\tp{\tilde{p}}
\def\tbmp{\tilde{\bm{p}}}
\def\tchi{\tilde{\chi}}
\def\tphi{\tilde{\phi}}
\def\nbar{\bar{n}}
\def\A{{\bf A}}
\def\B{{\bf B}}

\def\r{{\bf r}}
\def\k{{\bf k}}

\def\cO{{\cal O}}

\def\cE{{\cal E}}

\def\hM{\hat{M}}
\begin{document}

\title{\Large \bf Hidden order in dielectrics: string condensation, solitons, and the charge-vortex duality}
\author{\normalsize Sergei Khlebnikov \\
{\normalsize \it Department of Physics and Astronomy, Purdue University, West Lafayette, IN 47907, USA}}
\date{}
\maketitle

\begin{abstract}
Description of electrons in a dielectric as solitons of the polarization field 
requires that the interaction between the solitons 
(prior to their coupling to electromagnetism) is short-range.
We present an analytical study of the mechanism by which this is achieved. 
The mechanism is unusual in that it enables screening of electrically neutral soliton
cores by polarization charges. 
We also argue that 
the structure of the solitons allows them to be quantized as either fermions or bosons. 
At the quantum level, the theory has, in addition to the solitonic electric, elementary 
magnetic excitations, which give rise to a topological contribution to 
the magnetic susceptibility. 
\end{abstract}

\tableofcontents

\section{Introduction}

Recent years have brought attention to ordered 
phases of matter in which the order cannot be described within the
familiar paradigm of symmetry breaking and which therefore require complementary 
concepts.
Perhaps the most familiar example of such an order 
is provided by a type-II superconductor in
three dimensions: this system does 
not have a local order parameter, yet that does not preclude the existence of a continuous
phase transition \cite{Dasgupta&Halperin}.
By itself, then, the view that a hidden order, not associated with condensation 
of any local field, can exists in a condensed-matter system is hardly in doubt;
what is less well established is how common this phenomenon is.

In the example above, an important role is played by the topological defects---the vortex 
lines---and the fact that in the superconducting phase, where these lines can be described
semiclassically, the interaction between them is short-range. 
This is in contrast to long-range interactions mediated by the Nambu-Goldstone 
bosons in theories with spontaneously broken continuous global symmetries, as for example
in the Skyrme model \cite{Skyrme}, where the interaction energy 
of two solitons decays as $1/r^3$ at large distances \cite{Skyrme2}.

One may wonder if the property just discussed for the special case
of a superconductor, namely, the existence of topological defects (solitons)
with exponentially decaying interactions is in fact a general characteristic of 
a whole class of hidden orders. (We will continue to refer to these solitons as ``defects,"
despite the absence of a conventional long-range order, because they are
essentially screened versions of more conventional defects: monopoles or vortices.)

In the present paper, we consider from this perspective the description of electrons
and holes in a dielectric as solitons of the polarization field \cite{soli}. 
The existence of unconfined solitons in this model relies on the invariance of 
the static energy functional with respect to adding closed strings of quantized electric flux. 
This makes it similar in spirit to the models of string-net condensation, 
proposed as a description of a non-symmetry-breaking order in Refs.~\cite{Levin&Wen,Wen}.
Here, our focus is on developing an analytical argument that 
would allow us to establish the form of the soliton interactions.

We find that the interactions decay exponentially (prior to coupling of the solitons
to electromagnetism), as a consequence of a screening 
mechanism that allows an uncharged soliton core to source 
a polarization cloud with the total
charge of unity. The mechanism is distinct from (in a sense, even opposite to) 
the Debye screening in plasma, wherein a nonzero core charge produces a configuration 
of the overall charge zero. Rather, it is similar to the screening of 
the interaction between vortices in a superconductor by the vector field, where 
a defect in the order parameter sources a quantized magnetic flux. Indeed, we will
see that in two spatial dimensions, 
at the level of static configurations, the correspondence between the two mechanisms
becomes precise (a change of variables).
In three dimensions, that is no longer so; indeed, 
the defects in a superconductor and in an insulator even have different dimensionalities.
Nevertheless, some parallels remain.
A comparison of the two screening mechanisms in three dimensions
is presented in Table~\ref{table:comp}.

\begin{table}
\begin{tabular}{|c|c|c|}
\hline 
 & Magnetic case (superconductor) & Electric case (dielectric) \\
 \hline 
 Angular variable & phase $\theta$ & emergent gauge field $\bpsi$ \\
 \hline
 Screening field & vector $\A$ & scalar $\chi$ \\
 \hline
 Derivative object & covariant derivative & polarization vector \\
 & $\bmpi = \nabla \theta - \A$ & $\bmp = \nabla \times \bpsi + \nabla \chi$ \\
  \hline
 Quantized charge
 & magnetic flux = $\int \nabla \times \A$ & electric charge = $-\int \nabla^2 \chi$ \\
 \hline
\end{tabular}
\caption{\small Comparison of magnetic screening in a phase-only model of superconductor 
and electric screening in the solitonic theory of dielectrics in three dimensions. 
In either case, we focus on the screening of 
a topological defect in the corresponding angular variable due to the presence of another 
(``screening") field. }
\label{table:comp}
\end{table}

The similarity with the magnetic screening in superconductors becomes especially
close in the limit when
the size $\mu^{-1}$ of the polarization cloud of the soliton (the analog of the 
London depth in a superconductor) is large in lattice units.
We refer to this limit as an easily polarizable medium. In this case, the soliton
core, defined as the region at distances $r \ll \mu^{-1}$ from 
the soliton center, is practically uncharged. It is described \cite{soli} by 
a gauge theory---the lattice electrodynamics \cite{Wilson} of an emergent gauge 
field. At these distances, the soliton looks like a Dirac monopole \cite{Dirac}, 
a classical solution of that theory \cite{Polyakov}. 
The short-range interaction between the solitons is then seen
as a result of screening, over the scale $\mu^{-1}$, 
of the long-range fields of these monopoles by polarization charges.

The two-dimensional version of the present theory may be applicable to a synthetic 
dielectric: the insulating phase of an array of
Josephson junctions. (For an experimental
study of the phase diagram of such an array, see Ref.~\cite{Bottcher&al}.) In this
case, the role of the polarization field is played by the dipole moments of the
junctions, and the solitons are naturally interpreted as Cooper pairs.
For this to be feasible, one should be able to quantize the solitons as bosons.
Indeed, when we discuss the quantum version of the theory, we will observe
that the structure of the solitons allows them to be quantized as either 
fermions or bosons (or as anyons in two dimensions).

The paper is organized as follows. To make the presentation self-contained,
we begin, in Sec.~\ref{sec:sol}, with a review of the electron-as-soliton picture of 
Ref.~\cite{soli}. 
In Sec.~\ref{sec:screen}, we describe
an analytical calculation that allows one to understand the mechanism responsible
for screening the soliton interactions. 
We then proceed (in Secs.~\ref{sec:quant}--\ref{sec:obs}) to a discussion of quantum effects 
implied by the solitonic picture. 
The main outcome of this discussion is a curious version of the charge-vortex duality, 
one aspect of which is that quantization of vorticity, usually associated with superconductors, 
persists in dielectrics. Existence of quantized magnetic excitations
results in a topological contribution 
to the magnetic susceptibility, associated with closed-path tunneling of polarization charges.
We discuss prospects for its observability.
We summarize our results in Sec.~\ref{sec:concl}. Some details of the calculations for 
two spatial dimensions appear in the Appendix.

\section{Solitons in a nonlinear theory of dielectrics: a review}
\label{sec:sol}
\begin{figure}
\begin{center}
\includegraphics[width=1in]{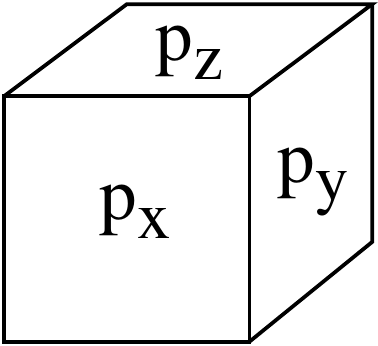}
\end{center}
\caption{\small 
A unit cell of a simple cubic lattice, with each face hosting a single component of the
polarization vector $\bmp$.}
\label{fig:cube}
\end{figure}

We begin with a review of the solitonic theory of electrons in a dielectric proposed 
in Ref.~\cite{soli}. Here, we focus on covalent crystals (such as silicon), for
which the polarization vector $\bmp$ is due entirely to the electrons. 
This vector is defined 
on a lattice, and its lattice divergence 
determines the polarization charge density $\rho$ by the usual formula
\be
\rho = - \nabla \cdot \bmp \, .
\label{rho}
\ee
In application to a specific material, the lattice may be expected
to correspond to that of the material, but for our present purpose (elucidation of the
screening mechanism) it is sufficient 
to consider a simple cubic lattice with unit spacing 
and the primitive vectors oriented along $x$, $y$, and $z$. The polarization vector 
$\bmp=(p_x,p_y,p_z)$ is defined
on the elementary faces (plaquettes) according to the following rule: 
a plaquette orthogonal to a primitive vector $\n$ hosts a single polarization component,
$\n \cdot \bmp$
(see Fig.~\ref{fig:cube}). We will assume that the components of $\bmp$, as well as the charge
density $\rho$, are measured in units
of $e/ 2\pi$, where $e$ is the electron charge. 

To describe the dynamics, we seek a Lagrangian that depends on $\bmp$ and its time
derivatives and is subject to an additional condition---invariance with respect to adding
closed $2\pi$ strings. Adding a $2\pi$ string is defined as selecting a 
directed path on the 
the dual lattice (formed by the centers of the unit cells of the original) and changing
the components of $\bmp$ at the plaquettes crossed by that path by $2\pi$. 
Physically, changing $\bmp$ by $2\pi$ at a plaquette corresponds to transporting 
a charge $e$ across that plaquette, 
as for instance when charges are separated through creation of
a particle-hole pair. Pulling the charges further apart creates an open 
$2\pi$ string between them.
The invariance with respect to adding closed strings means that the string has zero
tension. As a consequence, the energy of the charges does not grow with their separation.
This is precisely the rationale for requiring the invariance in question \cite{soli}, as
being able to separate electrons and holes 
to large distances without a persistent increase in energy
should be the case for any dielectric.

\subsection{Boundary conditions and the periodicity property}
\label{subsec:bc}

The invariance with respect to adding closed strings can be recast as a certain
periodicity condition
with the help of the Helmholtz 
decomposition of $\bmp$. The form of that decomposition is somewhat sensitive to the
boundary conditions. In this section and the next, we adopt those that allow us to 
consider isolated solitons (i.e., do not restrict the total charge of the configuration
to zero). They consist of the Neumann condition for the component $p_n$ of $\bmp$
orthogonal to the boundary ($b$) and the Dirichlet conditions for the components $\bmp_t$
tangent to it:
\be
\left. \part_n p_n \right|_b = \left. \bmp_t \right|_b = 0 \, .
\label{bc_vec}
\ee
For example, for those parts of the boundary that face the $x$ direction, we will use 
discretized versions of
\be
\left. \part_x p_x \right|_b = 0 \, , \hspace{3em} 
\left. p_y \right|_b = \left. p_z \right|_b= 0 \, .
\label{bc}
\ee
Physically, different boundary conditions correspond to different types 
of materials
that the sample can interface with. The conditions (\ref{bc}) allow the electric
current  to flow in and out of the sample
(the current density being proportional to $\part_t \bmp$) and so are 
appropriate for an interface with a superconductor. Conditions appropriate for an interface
with vacuum will be discussed in Sec.~\ref{sec:vort}.

As we will see shortly, the boundary conditions (\ref{bc_vec}) allow us to 
use the original Helmholtz decomposition,
\be
\bmp = \nabla \times \bpsi + \nabla \chi \, ,
\label{HH}
\ee
rather than Hodge's more general version. Here, $\bpsi$ is a field that lives on the edges
of the lattice (one component per edge, in accordance with the edge's direction), 
and $\chi$ is a scalar that lives on the sites of the dual lattice (centers of the unit 
cells). We use the continuum notation for the lattice derivatives. Thus, for example, the
lattice curl of $\bpsi$ is given by the circulation of $\bpsi$ around 
a plaquette:
\be
(\nabla \times \bpsi)_x(j,k,l) = \psi_z(j,k,l) - \psi_z(j, k-1,l) - \psi_y(j,k,l)
+ \psi_y(j,k,l-1) \, ,
\label{curl}
\ee
where triples of integers $(j,k,l)$ label both the plaquettes and the edges, and \eq{curl}
indicates how these two types of labels are related.
The discretization is chosen so that only $\chi$ contributes to the divergence in
(\ref{rho}): 
\be
\nabla \cdot \bmp = \nabla^2 \chi \, ,
\label{div}
\ee
where $\nabla^2$ is the lattice Laplacian.

To argue that (\ref{HH}) indeed holds, we first define $\chi$, for a given $\bmp$, 
as the solution to (\ref{div}) with the boundary condition $\left. \chi \right|_b= 0$.
We then have 
\be
\bmp = \bmg + \nabla \chi \, ,
\label{B}
\ee
where $\bmg$ obeys $\nabla \cdot \bmg = 0$ and the same boundary conditions as $\bmp$.
We now observe that it can be written 
as $\bmg = \nabla \times \bpsi$ for some $\bpsi$ obeying the boundary conditions
``dual'' to (\ref{bc_vec}):
\be
\left. \psi_n \right|_b = \left. \part_n \bpsi_t \right|_b = 0 \, .
\label{bc_psi}
\ee
Explicitly, the components of $\psi$ in a specific gauge are
\ba
\psi_x(x,y,z) & = & 
\int_0^z g_y (x,y,z') dz' - \int_0^y g_z(x,y',0) dy' \, , \label{psix} \\
\psi_y(x,y,z) & = & 
- \int_0^z g_x (x,y,z') dz' \, , \hspace{2em} \psi_z(x,y,z) = 0 \, , \label{psiyz} 
\ea
where the integration limits assume that three of the faces of the boundary coincide 
with the coordinate planes. The gauge arbitrariness inherent in the definition of
$\bpsi$ corresponds to transformations
$\bpsi \to \bpsi + \nabla f$ with an arbitrary function $f$ satisfying the 
Neumann boundary conditions. We can use this for instance to go over to the
gauge where $\nabla \cdot \bpsi = 0$.

Next, we observe that any closed $2\pi$ string 
can be obtained as a composition of elementary $2\pi$
strings, each encircling one edge of the lattice. Adding such an elementary string
amounts to changing $\bpsi$ on the edge by 
$2\pi$. Since this does not affect the value of the energy, one can set up an equivalence 
relation with respect to such changes. 
This turns $\bpsi$ into a compact gauge field---that
of the lattice electrodynamics of Ref.~\cite{Wilson}.

\subsection{Static energy functional}
\label{subsec:class}

At the classical level, one can start by looking at static configurations. For these, the
Lagrangian reduces to the negative of the static energy functional $E$:
\be
L[\bmp, \part_t \bmp] \to - \frac{1}{2\pi C} E[\bmp] \, ,
\label{LtoE}
\ee
where we have included a normalization coefficient in order to allow for arbitrary
normalization of $E$.

The simplest representative enough form of 
$E[\bmp]$, satisfying the periodicity requirement, is
\be
E[\bmp] = \sum_f V(\bmp) 
+ \half \sum_{cc'} (\nabla \cdot \bmp)_c M_{cc'} (\nabla \cdot \bmp)_{c'}  \, ,
\label{Eper}
\ee
where $V$ is a function that is $2\pi$ periodic in each of the components of 
$\bmp$, and $M_{cc'}$ are the elements of a symmetric matrix $\hM$.  
The first term here is a sum over the elementary faces (plaquettes). As we will see,
it represents finite polarizability of the medium. The second term is a sum over
the unit cells. In view of the relation (\ref{rho}), it can 
be interpreted as an intrinsic capacitive effect, with $\hM$ proportional 
to an inverse capacitance matrix. We assume $\hM$ to be positive definite.

Consider a long open $2\pi$ string originating at some point of the dual lattice
and extending to the boundary.
By virtue of \eq{rho},
the polarization charge density is nonzero only at the end of the string, with the total
charge accumulated there being $2\pi$, or $e$ in physical units.
Next, consider two opposite charges connected by a string. Since the string
tension is zero, the energy does not depend on the length of the string. We can say
that the interaction energy of the charges (prior to their coupling to  
electromagnetism) is zero. Due to the lattice derivative terms in $E$,
however, this configuration is not a classical solution:  
we expect those terms
to favor configurations where the charge density extends over some characteristic distance
near the end points. In this way, the ends of open strings become solitons. 
The interaction energy is now nonzero and, while it stands to reason that
it decays with the distance between the solitons, the
precise form of this decay remains to be determined. 

In the limit when the size of the soliton polarization cloud is large (in lattice units),
the form of the functional (\ref{Eper}) 
can be found by the derivative expansion. The first two terms are
\be
E[\bmp] = \sum_f V(\bmp) + \half \sum_c (\nabla \cdot \bmp)^2  \, .
\label{E}
\ee
We have used the arbitrariness of the coefficient in (\ref{LtoE}) to set the coefficient
of the second term here to one half. Then, $C$ in (\ref{LtoE}) can be thought of as the
self-capacitance of a unit cell, in appropriate units.

Throughout this paper, we assume cubic symmetry, meaning in particular that 
\be
V(\bmp) \approx \half \mu^2 \bmp^2  
\label{Vexp}
\ee
at small arguments. While the field $\bmp$ is not necessarily small at large
distances from the soliton, the field $\tbmp$, which is $\bmp$ with the string
subtracted, is. Because $V(\bmp)$ is periodic we can replace $\bmp$ in it
with $\tbmp$ and use the expansion (\ref{Vexp}) for $V(\tbmp)$.

Applicability of the derivative expansion requires $\mu \ll 1$ (in lattice units), 
the limit we refer to as an easily polarizable medium. 
In this case, the characteristic size of the polarization cloud is given by $\mu^{-1}$.
In principle,
there is no reason why this limit should hold for a particular dielectric.
Without relying on the derivative expansion, we would have to
consider the full linear operator corresponding to (\ref{Eper}) at small $\tbmp$,
\be
\hat{\cal L} = - \nabla^2 + \mu^2 \hM^{-1} \, .
\label{cL}
\ee 
As will be clear from the argument in Sec.~\ref{sec:screen}, the behavior of the
soliton fields at large distances is determined by that of the Green function
of (\ref{cL}).
In a system with translational invariance, this Green function can be written as
\be
G(\r) = \sum_{\k} \frac{1}{\lam_{\k}} e^{i\k\cdot\r} \, ,
\label{green}
\ee
where the sum is over the Brillouin zone (BZ), and $\lam_{\k}$ are the eigenvalues of 
(\ref{cL}). If we send the size of the lattice to infinity before taking the
large $r = |\r|$ limit, we can replace the sum in (\ref{green})
with an integral. Then, according to 
a well-known result of the Fourier series theory (see for example Sec.~12.D of 
Ref.~\cite{Arnold}), if we model $1/\lam_{\k}$ with a function analytic on the BZ,
the Green function (\ref{green}) decays exponentially at large $r$. In this manner,
we can convince ourselves 
that our argument for screening is unaffected by going from (\ref{E}) to the
more general expression (\ref{Eper}) (at least for a translationally invariant 
$\hat{M}$). 

For clarity of the presentation, in the discussion of the screening mechanism 
in Sec.~\ref{sec:screen} we then proceed with the simpler expression (\ref{E}).
We have used it also for numerical work, results of which are presented in 
Sec.~\ref{sec:num}, with
the parameter $\mu^{-1}$ there set typically just above the lattice scale 
($\mu^2 = 0.1$).
Discussion of observable effects in 
Sec.~\ref{sec:obs} does not rely on either of these choices. That discussion
is somewhat sensitive, though, to the ratio
\be
\alpha \equiv \mu^2 / C \, ,
\label{alpha}
\ee
in which $\mu$ appears in the Lagrangian (\ref{LtoE}). An estimate for this ratio
will be given shortly.

We wish to stress that the energy functional discussed so far represents the effect of
short-range interactions in the medium and, as such, applies prior to coupling of the
system to macroscopic electromagnetism.
The relation (\ref{rho}), however, allows us to immediately construct
an {\em additional} Hamiltonian describing the coupling to a static electric field. It reads
\be
H_{el} = - \frac{e}{2\pi} \sum_c (\nabla \cdot \bmp) \Phi_{el} \, ,
\label{elec}
\ee
where $\Phi_{el}$ is the electrostatic potential. 
(We will discuss the coupling to magnetic fields, which may be important
for nonstatic states, in Sec.~\ref{sec:vort}.) 

Consider the sample in a static uniform electric field $\cE = - \nabla \Phi_{el}$.
For weak fields, combining (\ref{elec}) with the expansion (\ref{Vexp}), we find
the equilibrium polarization to be
\be
\bmp = \frac{e \cE}{\alpha} \, ,
\label{unif}
\ee
where $\a$ is the parameter (\ref{alpha}).
In physical units, (\ref{unif}) corresponds to the polarization vector equal to
${\bf P} = e^2 \cE /(2\pi a \alpha)$, where $a$ is the lattice 
spacing. This allows one to relate the parameter $\alpha$ to the static dielectric 
constant $\eps$ of the material \cite{soli}:
\be
\alpha = \frac{2 e^2}{a (\eps - 1)} \, .
\label{aest}
\ee
For example, for silicon, using $a=5.4\, \mathring{\mathrm A}$ and 
$\eps = 12$ \cite{CRC}, we find $\a = 0.48$ eV.

Restoring the full form of $V(\bmp)$ will not be attempted here.
It affects the energy of an individual soliton but not the interaction of solitons 
at large distances. We have used the simplest periodic form
\be
V(\bmp) = \mu^2 [ 1 - \cos (\n \cdot \bmp ) ] 
\label{cos}
\ee
(the same as in Ref.~\cite{soli}) for the numerical work
but will not rely on this form in the discussion of observable effects 
in Sec.~\ref{sec:obs}.
 
\subsection{Restriction to two dimensions}
\label{sec:2D}

It is of interest to consider the restriction of the present theory to two
spatial dimensions. The counterpart of the energy functional (\ref{E}), 
with $\bmp$ now defined on the edges of a square lattice with unit spacing, is
\be
E = \sum_{\rm edges} V(\bmp) + \half \sum_c (\nabla\cdot\bmp)^2   \, .
\label{E2}
\ee
We expect this theory to apply to the insulating phase of an array of Josephson junctions,
with the meaning of the parameters being especially clear
in the case when the junctions are realized as short superconducting wires
(microbridges) connecting superconducting islands. 
For illustration, we will choose again $V(\bmp)$ in
the form (\ref{cos}), except that $\n$ is now the unit vector orthogonal to an edge.
The parameter $C$ in (\ref{LtoE}) then represents
the self-capacitance of a superconducting island, and the parameter
$\a$, \eq{alpha}, the fugacity of quantum phase slips in the wires 
\cite{Mooij&Nazarov,quantum_mech}.
(Solitons exist also in the one-dimensional version of the theory and, in that case,
directly in the continuum, where they are none other than the solitons of the 
sine-Gordon model. Here, however, our focus is on the two and three-dimensional cases.)

The analog of (\ref{HH}) in two dimensions is
\be
\bmp = \nabla \times \phi + \nabla \chi \, ,
\label{HH2}
\ee
where $\phi$ is a scalar defined on the vertices, and
$\nabla \times \phi = (\part_y \phi, -\part_x \phi)$. Adding an elementary closed string
surrounding a vertex corresponds to the change $\phi \to \phi \pm 2\pi$. Equivalence relation
with respect to such changes converts $\phi$ into an angular
variable. 

It follows from (\ref{HH2}) that the field $\bmpi$ orthogonal to $\bmp$, i.e., having 
components $\pi_i = - \eps_{ij} p^j$ (where $\eps_{xy} = - \eps_{yx} = 1$ and a sum over $j=x,y$ 
is implied), can be written as
\be
\bmpi = \nabla \phi - \nabla \times \chi \, .
\label{bmpi}
\ee
Referring to Table~\ref{table:comp}, we see that this field can be viewed as the
covariant derivative in the phase-only model of an equivalent 
superconductor, in the gauge where the
vector field $\A$ is a pure curl, $\A = \nabla \times \chi$. 
Upon substitution of $p^i = \eps^{ij} \pi_j$ into (\ref{E}),
the latter becomes precisely the static energy functional of such a superconductor. 
In particular, 
\be
(\nabla \cdot \bmp)^2 = (\nabla \times \A)^2 
\label{F2}
\ee
provides a Maxwell term for $\A$. The interaction
between vortices in this model is well known to decay exponentially \cite{LP}, 
an effect ordinarily attributed to screening of the interaction
by the vector field $\A$. In Appendix A, we interpret this effect directly in
terms of the scalar $\chi$, to reveal close parallels with screening of monopoles 
in the three-dimensional case.

\section{Numerical results}
\label{sec:num}

Before we proceed to discussion of the screening mechanism, we present 
results of numerical solution of the Euler-Lagrange equations corresponding to the
energy functional (\ref{E}) with the potential (\ref{cos}). These equations read
\be 
- \nabla (\nabla\cdot\bmp) + \part V / \part\bmp = 0 \, ,
\label{EL}
\ee
and should be satisfied on each plaquette of the cubic lattice.
The boundary conditions (\ref{bc}) allow configurations with a $2\pi$ string
extending from the center of any unit cell to the boundary of the lattice.

\begin{figure}
\begin{center}
\includegraphics[width=4in]{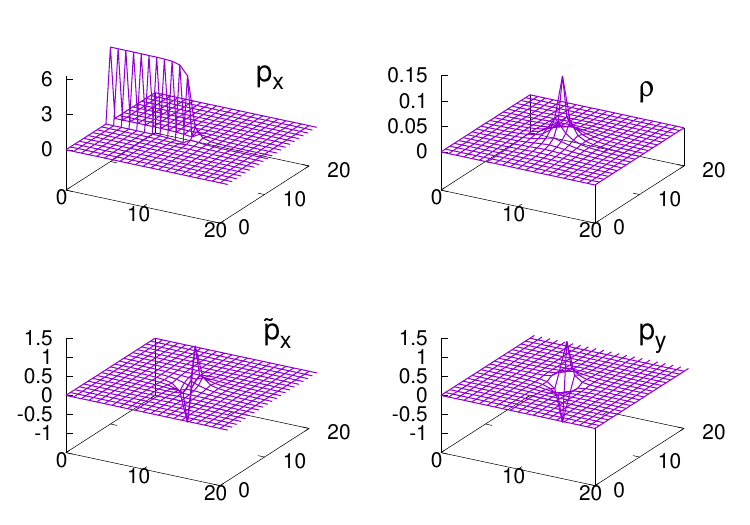}
\end{center}
\caption{\small
Profiles of the soliton field $\bmp$ and the density $\rho = - \nabla \cdot \bmp$ 
at the $(x,y)$ plane passing through the elementary ($q=1$) soliton center
in three dimensions, computed numerically on a $22^3$ grid for the potential (\ref{cos})
with $\mu^2 = 0.1$.}
\label{fig:sfields}
\end{figure}

In Fig.~\ref{fig:sfields}, we show results obtained by application of
the multidimensional 
Newton-Raphson method to \eq{EL}. The soliton is located at the center 
of the grid, with the $2 \pi$ string extending towards 
the negative $x$ direction. Note that, in addition to $p_x$, we plot the
component $\tp_x$ of the field
field $\tbmp$, obtained by subtracting the string from $\bmp$: in the continuum notation,
\ba
p_x(\r) & = & \tp_x(\r) + 2\pi \Theta(x_0 -x) \delta(y - y_0) \delta(z - z_0)  \, , 
\label{tpx} \\
p_y(\r) & = & \tp_y(\r) \, , \label{tpy} \\
p_z(\r) & = & \tp_z(\r) \, . \label{tpz}
\ea 
where $\r = (x,y,z)$, 
$\Theta$ is the step function, and $(x_0, y_0, z_0)$ is the location of the soliton. 

\begin{figure}
\begin{center}
\includegraphics[width=3.5in]{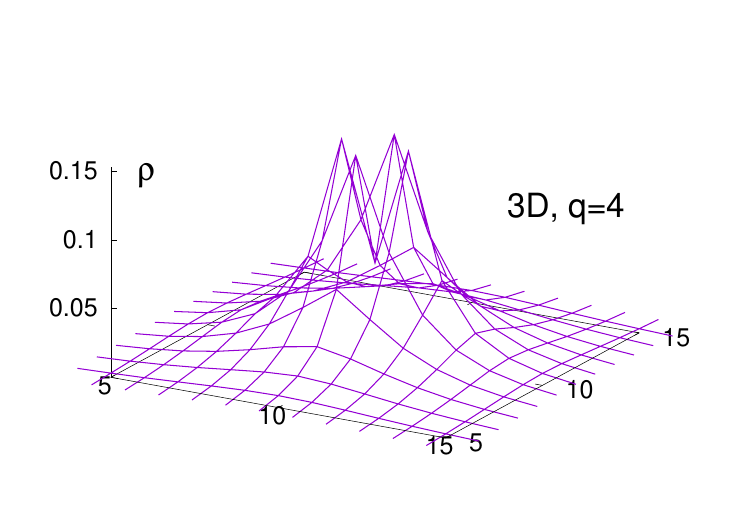}
\end{center}
\caption{\small
Same as in Fig.~\ref{fig:sfields} but for the density profile of the 
unstable solution with charge $q=4$. 
One can see four maxima of the density; there are two more off the plane, 
for the total of six.}
\label{fig:maxima}
\end{figure}

The story in two dimensions (2D) is very similar, with one exception. 
The latter concerns unstable solutions with integer 
multiples $q > 1$ of the elementary charge, by which we mean that they
each have a string carrying 
$2\pi q$ of the electric flux. In three dimensions (3D), 
we have found such solutions for all $q \leq 5$. The
ones with $q =4$ and 5 display a curious proliferation of maxima in the density profile, as
illustrated for the case $q = 4$
in Fig.~\ref{fig:maxima}. There are total of six density maxima in this case (two of these
lying off the plane
of the plot). That is so even though the components of $\bmp$ themselves do not show 
any intricate structure, being similar in the overall shape to those of the elementary 
(stable) soliton. In 2D, on the other hand, we have found an unstable solution (with a
single density maximum) for $q=2$ but none for larger $q$. 

\section{The screening mechanism}
\label{sec:screen}

We now describe an analytical method to compute the asymptotics of the fields and the soliton
interaction at large distances.
The strategy will be to assume (guided, say, by the numerical results of Sec.~\ref{sec:num})
that the field $\tilde{\bmp}$, obtained from $\bmp$ by subtracting the string, is small 
at large distances from the soliton center and argue that 
it is in fact small there {\em exponentially}.

Consider a single soliton centered at the origin with the string extending along the negative 
$x$ axis. We can absorb the
string by a redefinition of the field $\bpsi$ appearing in (\ref{HH}) into a new field
$\tbpsi$, as follows:
\be
\bpsi(\r) = \bm{\kappa}(\r) + \tbpsi(\r) \, ,
\label{tpsi}
\ee
where $\bm{\kappa}$ is the vector potential of a magnetic monopole. We will only
need the monopole
potential at large distances from the soliton center and so can use the expression for
it in the continuum, found by Dirac \cite{Dirac}; in the Cartesian components, it reads
\be
\kappa_x = 0 , \hspace{1em} 
\kappa_y = \frac{- z}{4 r^2 \cos^2 (\theta / 2)} ,  \hspace{1em}
\kappa_z = \frac{y}{4 r^2 \cos^2 (\theta / 2)} , 
\label{kappa}
\ee
where $r= |\r|$, and $\theta$ is the polar angle measured from the positive $x$ direction.
The tilded polarization vector, with the components given by (\ref{tpx})--(\ref{tpz}), is 
then
\be
\tbmp(\r) = \frac{\r}{2 r^3} + \nabla \times \tbpsi + \nabla \chi 
\equiv \bmb + \nabla \chi \, ,
\label{bmp3}
\ee
where we have defined a new field
$\bmb$, an expression for which can be read off the above.

Replacing $\bmp$ with $\tbmp$ does not change the periodic potential $V(\bmp)$. 
Upon that replacement, the energy functional (\ref{E}) becomes
\be
E = \half \sum_c (\nabla^2 \chi)^2 + \sum_f V(\tbmp) \, ,
\label{E3}
\ee
where we have used \eq{div}.
Consider the contribution to (\ref{E3}) from distances larger than some $R \gg 1$.
Assuming that the magnitude of  $\tbmp$ is already small there, we can apply
the expansion (\ref{Vexp}) to $V(\tbmp)$. Then, the contribution in question, 
in the continuum notation, is
\be
\Delta E_R \approx \! \half \int_{r > R} \hspace{-1em} d^3 x \left[ (\nabla^2 \chi)^2 
+ \mu^2 \tbmp^2 \right] ,
\label{ER}
\ee
where according to (\ref{bmp3})
\be
\tbmp^2 = (\nabla \chi)^2 + 2 \bmb \cdot \nabla \chi + \bm{b}^2 \, .
\label{tp2}
\ee
If we only want to use (\ref{ER}) as a basis for obtaining an equation for $\chi$ 
at large $r$, 
we can extend the integration there to all radii, but be prepared to encounter
a singularity at $r=0$ (where the expansion in small $\tbmp$ does not apply). 
The singularity appears when we integrate by parts the 
term with the dot product: this produces $\nabla \cdot \bmb$ which, according to the 
expression for $\bmb$ from (\ref{bmp3}), equals $2\pi \delta(\r)$. The equation 
for $\chi$ becomes
\be
(\nabla^2)^2 \chi - \mu^2 \nabla^2 \chi = 2\pi \mu^2 \delta(\r)  \, .
\label{eqchi}
\ee
The solution for $\nabla^2 \chi$, which by (\ref{div}) also gives the charge density,
is immediate:
\be
\nabla^2 \chi = - \frac{\mu^2}{2 r} e^{-\mu r} \, .
\label{dens}
\ee
The total charge is 
\be
Q = - \int \nabla^2 \chi d^3 x = 2 \pi \, ,
\label{Q}
\ee
independently of $\mu$. This corresponds to charge $e$ in physical units.

The form of the singularity on the right-hand side of (\ref{eqchi}) implies that,
while $\nabla^2 \chi$ is singular at $r=0$, $\chi$ itself is regular there. So, when
we solve \eq{dens} for $\chi$, we should include the $1/r$ solution to the homogeneous 
equation $\nabla^2 \chi = 0$ with the coefficient chosen so as to remove the singularity.
The result is
\be
\chi(r) = \frac{1}{2 r} (1 - e^{-\mu r}) \, .
\label{chi_sol3}
\ee
The same \eq{ER} can be used to see if the nonzero $\chi$ acts as a nontrivial source 
for $\tbpsi$.
We will see shortly, using a slightly different method, that it does not. Assembling all
the pieces of \eq{bmp3} together, we find that the solution for $\tbmp$ at the 
linearized level is
\be
\tbmp(\r) = \frac{\r}{2 r^3} + \nabla \chi = - \nabla \frac{e^{-\mu r}}{2 r} \, .
\label{tp_sol}
\ee
We see that the gradient of the long-range ($1/r$) part of $\chi$ precisely cancels 
the $\r / r^3$ term in 
(\ref{bmp3}). As a result, the fields $\tbmp$ decay exponentially.

It is useful to consider also another representation of the vector (\ref{bmp3}), namely,
\be
\tilde{\bmp} = \nabla \times \tbpsi + \nabla \tchi \, 
\label{repres}
\ee
where
\be
\tchi = \chi - \frac{1}{2r} \, .
\label{tchi}
\ee
Using (\ref{repres}) in (\ref{ER}) and extending, as before, the
integration to all radii, we obtain 
\be
\Delta E \approx \! \int  d^3 x \left\{ 
\half [ \nabla^2 \tchi - 2\pi \delta(\r)]^2 
+ \frac{\mu^2}{2} \left[
( \nabla \tchi)^2  + (\nabla \times \tbpsi)^2 \right] \right\}  .
\label{Etchi}
\ee
The variational problem for $\tchi$ thus amounts to finding a solution to
\be
\nabla^2 [ \nabla^2 \tchi - 2\pi \delta(\r) ] - \mu^2 \nabla^2 \tchi = 0 \, .
\label{eq_tchi}
\ee
This is satisfied if we let $\nabla^2 \tchi - 2\pi \delta(\r) = \mu^2 \tchi$,
meaning that the solution is none other than the suitably normalized Green function 
of the Helmholtz equation,
\be
\tchi(\r) = - \frac{e^{-\mu r}}{2 r} \, .
\label{GF}
\ee 
Combining this with \eq{tchi}, we obtain again the solution (\ref{chi_sol3}) for $\chi$. 
This way of looking at the problem, however, underscores the generality 
of the screening mechanism. Indeed, we have already noted (in Sec.~\ref{subsec:class}),
that the argument for screening is unaffected if we replace (\ref{E}) with the more general 
expression (\ref{Eper}). We can now see that by noting that this replacement modifies 
the first term in (\ref{eq_tchi}) in such a way
that, instead of (\ref{GF}), the solution is now proportional to the more general
Green function (\ref{green}).
We also note that,
according to (\ref{Etchi}), the field $\tbpsi$ is not sourced at the linear level: 
the variational
equation for it is $\nabla\times(\nabla\times \tbpsi) = 0$, 
and the solution, in a suitable gauge, is $\tbpsi = 0$.

We find this screening effect remarkable, as it is quite distinct from the usual (Debye)
screening of an external charge in a plasma. In that case, the induced charge density
(the counterpart of our $\rho$) is $\rho_{ind} = - \frac{\mu^2}{4\pi} \Phi_{el}$, 
where $\Phi_{el}$ is the
electrostatic potential, and $\mu^2$ is a coefficient depending on the parameters of 
the plasma. The Poisson equation $-\nabla^2 \Phi_{el} = 4\pi (\rho_{ext} + \rho_{ind})$ becomes 
\be
- \nabla^2 \Phi_{el} + \mu^2 \Phi_{el} = 4\pi \rho_{ext} \, .
\label{ext}
\ee
For $\rho_{ext}$ corresponding to a unit point charge, the solution is
\be
\Phi_{el}(\r) = \frac{e^{-\mu r}}{r} \, .
\label{ext_sol}
\ee
This solution is clearly different from (\ref{chi_sol3}).
In particular, the total charge contained in it is zero, while that in (\ref{chi_sol3}) 
is not. 

The physical pictures of the two effects are also completely different. 
This is most easily seen in the
limit of a highly polarizable medium, when $\mu \ll 1$. Then, in the core region
$1 \ll r \ll \mu^{-1}$, 
(\ref{ext_sol}) is the potential of a point charge, surrounded at larger distances
by a polarization cloud. The solution (\ref{chi_sol3}), on the other hand,
is nearly constant at the core, so the core is nearly neutral. 
The fields $\bmp$ there are essentially those of the monopole of $\bpsi$.
The entire charge in this case is contained in a cloud
of radius $r \sim \mu^{-1}$, apparently sourced by that neutral core. 

On the other hand, one may note an analogy between the present mechanism and the screening
of vortex lines
in extreme type-II superconductors, where a nonmagnetic core sources magnetic flux.
In this respect,
the main result of this section is that a closely parallel mechanism exists for point-like
(as opposed to line-like) defects in three dimensions.

The field $\tbmp$ and the charge density $\rho$ (the only objects on which the energy
functional depends) decaying exponentially for a single
soliton implies that so does also the soliton-soliton interaction. An explicit expression 
for it can be obtained in the limit
$\mu \ll 1$. In this case,
for a soliton and an antisoliton separated by a distance $L_s \gg 1$, the interaction 
energy accumulates mostly at distances $r \gg 1$ from the soliton centers, where
we can use the quadratic approximation for $V(\tbmp)$ and work directly in the continuum.
The result is
\be
E_{int} = \int d^3 x \left[ (\nabla^2 \chi)_1 (\nabla^2 \chi)_2 + \mu^2 \tbmp_1 \cdot \tbmp_2 
\right] \, ,
\label{Eint}
\ee
where the subscripts 1 and 2 refer to the soliton and antisoliton, respectively. 
A direct computation gives
\be
E_{int} = - \pi \mu^2 \frac{e^{-\mu L_s}}{L_s} \, .
\label{Eint3}
\ee
For comparison, the energy of a single soliton (which always has to be computed on the lattice)
is approximately linear in $\mu^2$ at small $\mu$, $E_{sol} \approx 4.68 \mu^2$, but
deviates down from this line at larger $\mu$.
For $\mu^2 = 0.1$, $E_{sol} = 0.42$. 
Upon including the the
normalization factor from (\ref{LtoE}), the classical energy of the soliton becomes
\be
E_{sol}' = E_{sol} / (2\pi C) \, ,
\label{Esol'}
\ee
where $E_{sol}$ is the dimensionless energy discussed above.

\section{Towards a quantum theory}
\label{sec:quant}

So far, our considerations have been entirely at the classical level. To proceed to 
a quantum theory, we
must include the static energy $E$, given by (\ref{E}) or (\ref{E2}), 
as a part of a Lagrangian supplying the time
dependence. The simplest choice, quadratic in time derivatives, of the Lagrangian in 3D is
\be
L = \frac{1}{4 \pi K} \sum_f (\part_t \bmp)^2 - \frac{1}{2\pi C} E \, ,
\label{lagr}
\ee
where the sum is over the elementary faces, and $K$ and $C$ are constant coefficients. 
In the first term, $\bmp$ refers to the single component that the polarization vector has
at each face. Note that
$\part_t \bmp$ is the polarization current density (in units of $e/2\pi$), so $K$ can
be interpreted as the intrinsic inverse self-inductance per unit face, in appropriate units.
[More generally, one can replace the first term with one including an inverse inductance 
matrix, similar to the
inverse capacitance matrix $\hat{M}$ in (\ref{Eper}).]

In 2D, the sum in (\ref{lagr}) has to be replaced by a sum over the edges. We note that
in this case the first term in (\ref{lagr}) breaks the equivalence of our model with 
a phase-only superconductor, based on the transition from (\ref{HH2}) 
to (\ref{bmpi}). As we will see, the model exhibits instead a curious version of self
duality, with both electric and magnetic excitations present.

In what follows we set $\hbar =1$, so that
$K$ has the dimension of energy. Since we consider the case of purely electronic polarization, 
we expect $K$ to be of order 1 eV, i.e., the same order as the parameter $\a$ defined
in (\ref{alpha}).

When there is a finite density of solitons, effects of their
quantum statistics can become important.
In Fig.~\ref{fig:sfields}, we see that, while the charge density apparently respects
the cubic symmetry of the model, the individual polarization components do not. 
This means that the rotational configuration space
of the soliton is the full SO(3) in 3D or SO(2) in 2D. We note that the lattice 
pins the orientation of the
soliton relative to the lattice directions, just as it pins its position 
to the center of a unit cell.
This pinning, however, is only an energy barrier and does not affect the topology of 
the configuration
space. Following the well-known argument \cite{FR}, we then conclude that the 
solitons can be quantized either as bosons or fermions 
in 3D, or as anyons in 2D. 

Conventional dielectrics will require quantization of the solitons as fermions.
Quantization as bosons may be of interest in the context of a synthetic dielectric:
an array of Josephson junctions in its
insulating phase (see Ref.~\cite{Bottcher&al} for a study of the phase diagram of such 
an array). In this case, the natural physical unit of charge, with respect to 
which we should require periodicity is $2e$ rather than $e$. Solitons of charge $2e$ 
can be interpreted as Cooper pairs.

Observation of a charge-$2e$ soliton in an array of Josephson junctions would be
of interest in its own right and would also suggest that these systems may
be useful for understanding the physics of 
conventional dielectrics, offering a degree of control
over the parameters unavailable in the latter case. 
In what follows, though, we focus on novel magnetic 
excitations, which do not activate the charge subsystem at all and are thus
insensitive to the statistics of the solitons.

\section{Magnetic vortices in dielectrics}
\label{sec:vort}

At the quantum level, the theory (\ref{lagr}) has also another type of excitation, 
one we have not 
discussed so far. This is the magnetic vortex. It is identified most readily by looking 
first at the 2D version. Let us use for that
the Helmholtz decomposition (\ref{HH2}) and proceed to the canonical quantization of 
the degree of freedom represented by $\phi$.
The canonical momentum conjugate to $\phi$ is
\be
\Pi = (2 \pi K)^{-1} (- \nabla^2) \part_t \phi \, .
\label{Pi}
\ee
Note that, just as $\phi$, $\Pi$ is defined on the vertices (sites) of the lattice.
For reasons that will become clear shortly, we refer to it as vorticity.
Because $\phi_v$ (on each lattice site) is a $2\pi$-periodic angular variable, each $\Pi_v$ is 
quantized in integer units.

The boundary condition for $\phi$ corresponding to (\ref{bc_vec}) is
$\left. \part_n \phi \right|_b = 0$, where $\part_n$ is the normal derivative.
Eq.~(\ref{Pi}) then implies a constraint (a
global equivalent of the Gauss law) $\sum_v \Pi_v = 0$, where the sum is over all the vertices.
As already discussed in Sec.~\ref{subsec:bc}, the boundary condition (\ref{bc_vec}) allows for
a flow of charge into the sample;
it thus allows consideration of individual solitons. The above constraint, though, indicates
that (\ref{bc_vec}) does not allow 
consideration of individual quanta of vorticity (vortices). This may be understood by
noting that it prohibits
a nonzero density $\part_t \bmp_t$ of the tangential electric current at the boundary, 
required to support an individual vortex.
For this reason, it will be convenient for us to adopt in this section the boundary
condition dual to the above, i.e.,
\be
\left. \phi \right|_b = 0 \, .
\label{new_bc}
\ee
\eq{new_bc} prevents an inflow of charge but allows an inflow of vorticity.
As such, this boundary condition is appropriate at an interface of the sample with vacuum.
One can see that $\sum_v \Pi_v$ is now unconstrained.

Inverting (\ref{Pi}) to find the current density $\nabla \times \part_t \phi$ 
carried by $\phi$, 
we see that the eigenstate
of $\Pi_v$ that has $\Pi_v = \delta_{v,v_0}$, i.e., $\Pi_v = 1$ at some $v = v_0$ 
and zero everywhere else,
describes a vortex pattern of current around $v_0$.  
We will refer to these vortices as 
magnetic, to distinguish them from the vortices at the cores of the electrically 
charged solitons
considered in the preceding sections.
 
The kinetic term for $\phi$ in the Hamiltonian corresponding to (\ref{lagr}) is
\be
H_{kin} = \pi K \sum_{v} \Pi (- \nabla^2)^{-1} \Pi \, .
\label{Hkin}
\ee
To understand its role, it is useful to consider first the limit when $K$ is the
largest energy scale in the problem, namely, $K \gg \a$. 
As discussed above, we do not expect this to be the case in a conventional covalent 
dielectric.
Instead, as we will now argue, this limit bears hallmarks of a superfluid state.
We may then expect it to 
be realizable in an array of Josephson junctions, where the phase diagram is
known to include
such a state (see, for example, Ref.~\cite{Bottcher&al}). 

To argue the connection with superfluidity, we
note that, in the limit $K \gg \a$, the term (\ref{Hkin}) is the leading term 
in the Hamiltonian of $\phi$, so the common
eigenstates of all $\Pi_v$ are also approximate eigenstates of the Hamiltonian. 
Substituting 
\be
\Pi_v = \delta_{v,v_1} - \delta_{v,v_2} 
\label{pair}
\ee
in (\ref{Hkin}) we find that this term describes a 
logarithmic interaction between the magnetic 
vortices. This makes
it clear that the vortices in question are none other 
than the usual vortices of a discrete 2D superfluid, with $K$ being proportional to the
Josephson energy.

As one moves across the phase boundary to the insulating phase of the array, where 
$K \ll \a$, one can no longer rely solely on
$H_{kin}$, as the second term in (\ref{lagr}) becomes important. In this case,
one may prefer to think of charges as solitons of the theory identical to the one described in
Sec.~\ref{sec:sol}, except possibly with a different value of the minimal charge (cf. the
discussion in Sec.~\ref{sec:quant}).
On the other hand, the magnetic vortices can no longer be described semiclassically. 
We should use instead the fully quantum definition,
with the addition of
a vortex at site $v_0$ defined as the action of any operator $\cO(v_0)$ such that
\be
[\Pi_v, \cO(v_0)] = \delta_{v, v_0} \cO(v_0) \, ,
\label{cO}
\ee
where the square brackets denote a commutator.

Another way to arrive at the interpretation of the eigenstates of $\Pi$ as magnetic vortices
is to consider interaction of our system with
a vector potential $\A$. It is given, in parallel to (\ref{elec}), by the Lagrangian
\be
L_{mag} = \frac{e}{2\pi c} \sum \A \cdot \part_t \bmp \, ,
\label{mag}
\ee
where $c$ is the speed of light in vacuum, and the sum is over the plaquettes in 3D or the
edges in 2D.
Addition of (\ref{mag}) changes the canonical 
momentum of $\phi$ from (\ref{Pi}) to
\be
\Pi = (2 \pi K)^{-1} (- \nabla^2) \part_t \phi + (e / 2\pi c) B \, ,
\label{Pi2}
\ee
where $B = (\nabla \times \A)_\perp$ is the perpendicular magnetic field 
in units where the lattice spacing is set to one. This leads
to the replacement $\Pi \to \Pi - (e / 2\pi c) B$ in the kinetic
term (\ref{Hkin}), which shows that a state with a nonzero expectation value of $\Pi_v$
carries a density of the magnetic moment.

In 3D, the counterpart of $\phi$ is the gauge field $\bpsi$ of (\ref{HH}). 
Instead of the global symmetry
$\phi \to \phi + \mbox{const}$, we now have the gauge symmetry $\bpsi \to \bpsi + \nabla f$. 
Accordingly, the constraint is now the usual Gauss law $\nabla \cdot \bm{\Pi} = 0$, where
\be
\bm{\Pi} = \frac{1}{2\pi K} \nabla \times (\nabla \times \part_t \bpsi) 
+ \frac{e}{2\pi c} \B
\label{Pi3}
\ee
is the momentum conjugate to $\bpsi$; just as $\bpsi$ itself, it lives on the lattice
edges. The second term here, with $\B = \nabla \times \A$ is due to the coupling (\ref{mag}).
As usual in lattice gauge theories, the Gauss law represents conservation of flux 
at the vertices \cite{Kogut&Susskind}, so the magnetic vortices become vortex lines.

\section{Prospects for observability}
\label{sec:obs}

Quantization of vorticity in a dielectric is a somewhat unusual prediction
of the present theory, so it is natural to ask what may be its observable consequences.
We start by considering magnetic fluctuations in the ground state of a dielectric crystal. 
The simplest such fluctuation is spontaneous 
formation and disappearance of a localized magnetic 
moment. We will describe this process using a Euclidean classical solution (instanton) of 
the theory (\ref{lagr}). The solution is purely magnetic, meaning that the charge
field $\chi$ on it is zero. Upon continuation to the Euclidean time
$\tau = i t$, the equations of motion become
\be
\part_\tau^2 \bmp - \frac{K}{C} \dpar{V}{\bmp} =  0 \, , \hspace{3em}
\nabla \cdot \bmp = 0 \, .
\label{eucl} 
\ee
Note that the first of these is a separate 
equation for the individual $\bmp$ on each plaquette. These $\bmp$, however, 
are connected by the second equation. 

By virtue of the charge neutrality condition $\nabla \cdot \bmp = 0$, the capacitive 
terms, either those of the matrix type as in (\ref{Eper}) or the simplified one as
in (\ref{E}), do not contribute to (\ref{eucl}). Accordingly, the parameters $\mu^2$
and $C$ appear in (\ref{eucl}) only through their ratio, $\a$, an estimate for which was given
in Sec.~\ref{subsec:class}.

We may use \eq{eucl} without any changes also in the case when the crystal is placed
in a time-independent magnetic field. The reason is that, for such a field,
the additional Lagrangian (\ref{mag}) is ``topological,'' 
in the sense that, being a perfect time derivative, it does not contribute to 
the classical equations of motion. At the quantum level, however,
it can contribute a phase factor to a transition amplitude.

In a single instanton, one of the components of $\bpsi$, say, $\psi_z$, changes
from 0 to  $2\pi$ at a single edge during the Euclidean
time interval $-\beta/2 < \tau < \beta/2$, 
where $\beta$ is a large parameter, while remaining zero at all the other edges. 
The corresponding $\bmp = \nabla \times \bpsi$ is then nonzero 
only at the four plaquettes that meet at the edge in question. 
Thus, the instanton corresponds to a circular current, resulting from tunneling 
of polarization charge around that edge.
One can verify that the charge neutrality condition 
$\nabla \cdot \bmp = 0$ is satisfied. 

The Euclidean action of the instanton is the sum of four individual contributions,
one from each of the four plaquettes in question. Two of these have $p_x = \pm \psi_z$, 
and the other two $p_y = \pm \psi_z$ (recall that there is only one component of
$\bmp$ per plaquette). A potential
$V(\bmp)$ respecting the cubic symmetry has the same value, $V(\psi_z)$, 
on all four. Combining this with the contribution from the kinetic term in (\ref{lagr})
and including the interaction (\ref{mag}), where $\A$ corresponds to a 
constant uniform magnetic field $B$ in the $z$ direction, we find the Euclidean
action to be
\be
S_E = \frac{1}{2\pi} \int d\tau \left\{ \frac{4}{2K} (\part_\tau \psi_z)^2 
+ \frac{4}{C} V(\psi_z)  - \frac{i e B a^2}{c}  \part_\tau \psi_z \right\} \, .
\label{SE}
\ee
We have restored the lattice spacing $a$, so that $B$ is now in physical units.

One may note that the action (\ref{SE}) is formally
identical to that of a short superconducting wire written 
in terms of the variable---the dipole moment---dual to the gradient of the phase of the
order parameter between the wire's ends \cite{Mooij&Nazarov,quantum_mech}. 
In the present system, the equivalent ``wire'' is
formed by the current circulating around a lattice edge.

The profile $\psi_z(\tau)$ of the instanton solution is determined by the Euler-Lagrange
equation corresponding to (\ref{SE}), 
\be
\part_\tau^2 \psi_z + (K / C) V'(\psi_z) = 0
\label{eq_inst}
\ee
with the boundary conditions $\psi_z = 0, 2\pi$ at $\tau \to \pm \beta/2$ respectively.
(As already noted, the topological term, proportional to $B$, does not contribute
to the equation of motion.)

An antiinstanton is obtained from the instanton by reversing the direction of the current.
This only affects the last term in (\ref{SE}). The integration over $\tau$ in that term
can be trivially carried out, so the Euclidean action can be written as
\be
S_E = S_0 \mp \frac{i e \Phi}{c} \, ,
\label{Phi}
\ee
where the upper (lower) sign corresponds to an instanton (antiinstanton), $\Phi = B a^2$,
and $S_0$ is the real part of the action. In the ground-state partition function, given
by the path integral of $e^{-S_E}$, each instanton thus contributes a phase 
factor $e^{i e \Phi / c}$, and
each antiinstanton a factor $e^{-i e \Phi / c}$. This results in the ground-state energy,
$E_{gs}(B)$, becoming $B$-dependent. 

If the instantons are relatively rare,
their contribution to the path integral can be computed in the dilute-gas
approximation. It leads to appearance in the partition sum of the factor
\be
Z_{inst} = \sum_{n_+ = 0}^{\infty} \frac{(\nbar N_{tot} \beta)^{n_+}}{n_+!} e^{i e \Phi n_+ / c} 
= \exp \left( \nbar N_{tot} \beta e^{i e \Phi / c} \right) \, ,
\label{Zinst}
\ee
where $\nbar$ is the instanton density (the average number of instantons
per edge per unit time), $N_{tot}$ is the
total number of edges in the $z$ direction, and $\beta$ is the total Euclidean
time duration. An individual term in the sum is the contribution from paths
containing $n_+$ instantons. Antiinstantons contribute a similar factor but with the
sign in front of $\Phi$ reversed. The product of the two factors,
\be
Z_{inst} Z_{anti} = \exp \left[ 2 \nbar N_{tot} \beta \cos(e \Phi / c)  \right] \, ,
\label{prod}
\ee
thus includes contributions of paths with arbitrary numbers of instantons and 
antiinstantons. 
Writing the partition function as $\exp(-\beta E_{gs})$, we obtain 
\be
E_{gs}(B) = - 2 \nbar N_{tot}  \cos (e B a^2 /c) + {\rm const} 
\label{E0}
\ee
for the ground-state energy. If the volume magnetic susceptibility $\chi_v$ is
small, it can be obtained by 
differentiating $-E_{gs}$ twice with respect to $B$ at $B = 0$ and dividing the result 
by $N_{tot} a^3$:
\be
\chi_v = - \frac{2 \nbar e^2 a}{c^2} = - 7.40 \times 10^{-6} 
(\nbar / {\rm eV}) (a / \mathring{\mathrm A})
\label{chiv}
\ee
(in CGS units). 

For a large $S_0$, the instanton density $\nbar$ is suppressed by the semiclassical
factor $e^{-S_0}$.
In the case of the cosine potential (\ref{cos}), 
the solution to (\ref{eq_inst})
is the soliton of the sine-Gordon model, and
\be
S_0 = \frac{16}{\pi} (\a / K)^{1/2} \, .
\label{S0}
\ee
We see that for $\a \sim K$, the suppression by $e^{-S_0}$ is at most moderate.
One should keep in mind that 
the value of the instanton action depends
on the precise shape of the potential $V(\bmp)$. Since (\ref{cos}) is only a model chosen 
for illustration, the value (\ref{S0}) may in practice not be accurate enough. 
Still, one may wonder if, with suitable modeling, Eq.~(\ref{chiv}) can be a useful way
to represent the susceptibility of a covalent crystal.
 
In addition, it is known that tunneling amplitudes often grow rapidly 
(exponentially) with the amount of excitation available in the initial state. 
To prevent that initial excitation from dispersing, it appears advantageous to consider, 
instead of a large uniform sample, an array of small dielectrics (nanocrystals). 
Enhancement of the magnetic response of small dielectrics is well known in optics 
(for a review, see Ref.~\cite{Jahani&Jacob}), where it is described as a resonance 
of Mie’s classical scattering theory. 
It would be interesting to see if the enhancement of the
topological dc susceptibility (\ref{chiv}) at such a resonance can be detected 
experimentally. 
One may notice that the use of small linear dimensions of a sample for 
control of the magnetic
response in this proposal
has some similarity with the use of a small superconductor---a Cooper-pair 
box \cite{Buttiker,Bouchiat&al}---for control of the electric charge. 

\section{Conclusion}
\label{sec:concl}

We thus arrive at a curiously complete instance of the charge-vortex duality, with a near 
perfect symmetry between the electric and magnetic charges. Each can be considered, 
under suitable conditions, either as elementary or as solitonic. This duality is 
characteristic of theories that obey a certain periodicity requirement---namely,
the invariance of the static energy with respect to adding closed strings of integer 
electric flux.
In the present paper, we have focused on the consequences of this periodicity 
for dielectrics, where two effects stand out. 

One is the existence of solitonic electrically charged excitations, with a charge density 
possibly extending over several lattice volumes \cite{soli}. 
Here, we have developed an analytical understanding of the mechanism by which the total electric
charge of unity is accumulated
over a finite distance, starting from an uncharged core. (This is opposite
to the usual Debye screening, which starts with a charged particle and results in an object of the overall 
charge zero.) We have obtained analytical results for the interaction between the solitons,
confirming that it is short-range (prior to coupling
to electromagnetism).
We have also argued that the structure of the solitons allows them to be quantized as either 
fermions or bosons (or anyons in 2D).

The second effect is related to quantization of vorticity. It amounts to
a topological contribution to the magnetic susceptibility, 
associated with closed-path tunneling of polarization charges.
A possible route to enhancing it may be access to an excited state of a small crystal 
via a Mie resonance.

Finally, while 
experimental discovery of either effect in a conventional dielectric would be most interesting, 
one may also consider searching for them in a synthetic insulator, formed by an array of 
Josephson junctions. In that case, one may be able adjust the
parameters to one's advantage. 

\appendix
\section{Screening in two dimensions}

As we already noted, in two dimensions, the change of variables
from $\bmp$ to $\bmpi$, \eq{bmpi}, allows one to understand screening of the interaction
between solitons 
in the present model as that between vortices in an equivalent superconductor, with the
vector field of the latter
given by $\A = \nabla \times \chi$. In this Appendix, we consider  
the screening
mechanism in 2D using instead 
the same method as used in the main text for 3D, to underscore the
parallels between the two cases.

In 3D, the key ingredient was the definition (\ref{tpsi}) of the field $\tbpsi$ that
absorbs the vector potential of a monopole.
In 2D, the Helmholtz decomposition is
given by (\ref{HH2}), and the analog of $\tbpsi$ is the field $\tphi$ defined,
in the continuum notation, by
\be
\phi(x,y) = \arctan (y / x) + \tphi(x,y) \, .
\label{tphi}
\ee
The first term is a vortex that carries a $2\pi$ string in the negative $x$ direction.
The boundary conditions (\ref{bc_vec}) correspond to $\left. \part_n \phi \right|_b = 0$
and the same for $\tphi$. 

The counterpart of (\ref{bmp3}) is 
\be
\tbmp(\r) = \frac{\r}{r^2} + \nabla \times \tphi + \nabla \chi 
\label{bmp2}
\ee
for the field $\tbmp$, which is $\bmp$ with the string subtracted. Here $r = |\r|$
and $\nabla \times \tphi = (\part_y \tphi, -\part_x \tphi)$. 
The remainder of the argument of Sec.~\ref{sec:screen} is now 
transferred to the 2D case almost literally. The equation for $\chi$ at large $r$ 
is the same \eq{eqchi}, except that the delta
function is two-dimensional. The solution, determined up to a constant, $c$, is
\be
\chi(r) = - K_0(\mu r) - \ln r + c \, ,
\label{chi_sol2}
\ee 
where $K_0$ is the modified Bessel function of order zero. The charge density is
\be
- \nabla^2 \chi = \mu^2 K_0 (\mu r) \, ,
\label{dens2}
\ee
corresponding to the total charge of $2\pi$.

To rewrite (\ref{bmp2}) similarly to (\ref{repres}), i.e., as 
\be
\tbmp = \nabla \times \tphi + \nabla \tchi \, ,
\label{repres2}
\ee
we define 
\be
\tchi = \chi + \ln r - c  \, .
\label{tchi2}
\ee
The solution (\ref{chi_sol2}) corresponds to an exponentially decaying
$\tchi = - K_0(\mu r)$. 
An expression for the energy analogous to (\ref{Etchi}) is
\be
\Delta E \approx \! \int  d^2 x \left\{ 
\half [ \nabla^2 \tchi - 2\pi \delta(\r)]^2 
+ \frac{\mu^2}{2} \left[
( \nabla \tchi)^2  + (\nabla \times \tphi)^2 \right] \right\}  .
\label{Etchi2}
\ee
It can be used to show that $\tphi$ is not sourced at the linear level. 
Indeed, the equation for $\tphi$ following from it is $\nabla^2 \tphi = 0$,
and the solution, after a suitable fixing of 
the inessential additive constant, is $\tphi = 0$. Combining this with the result
for $\tchi$, we see that the fields $\tbmp$ decay exponentially.

In Sec.~\ref{sec:screen}, we contrasted the screening mechanism there
with the Debye screening
in a plasma. Here, a natural point of comparison is polarization caused
by an external point charge in a linear 
2D capacitive environment. An example of such an environment is an array of tunnel
junctions, such as considered for example in Ref.~\cite{Mooij&al}.
The role comparable to $\chi$ is played in that case by the electrostatic potential
$\Phi_{el}$, which obeys an equation analogous to (\ref{ext}). 
Unlike the flux of $\nabla \chi$, however, 
the flux of $\nabla \Phi_{el}$ though a circle at infinity is zero.
One way to express the difference with the present case is to observe
that, in the model of Ref.~\cite{Mooij&al}, $\Phi_{el}$ is the only dynamical variable.
Here, on the other hand, we also have the angular variable $\phi$, capable of supporting
point-like defects (vortices). In other words, the full set of variables in our model is
a vector, $\bmp$, not a scalar.

\begin{figure}
\begin{center}
\includegraphics[width=3.5in]{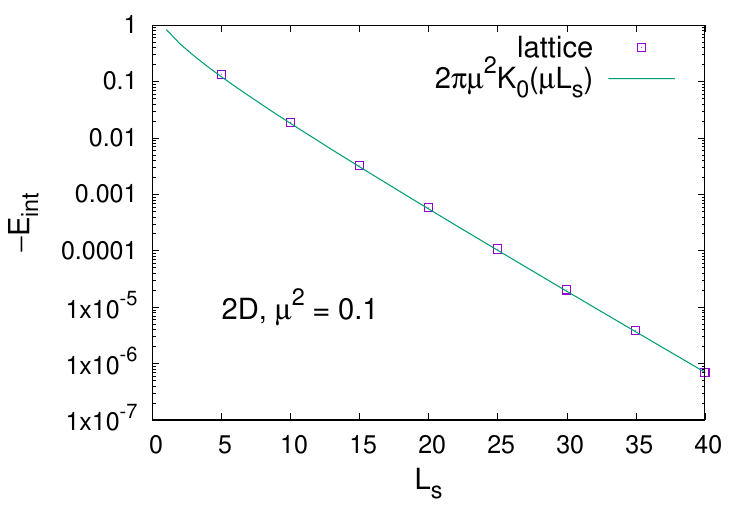}
\end{center}
\caption{\small Absolute value of the interaction energy of a soliton-antisoliton pair in 2D as 
a function of the separation, computed on a $100\times 50$ grid for $\mu^2 = 0.1$. $K_0$
is the modified Bessel function appearing in the analytical result (\ref{Eint2}).}
\label{fig:Eint}
\end{figure}

In the limit  $\mu \ll 1$, the interaction between a soliton and an antisoliton separated 
by a distance $L_s \gg 1$,
can again be obtained from the continuum expression (\ref{Eint}), except that 
the integral is 
now two-dimensional. The result is
\be
E_{int} = - 2\pi \mu^2 K_0(\mu L_s) \, ,
\label{Eint2}
\ee
the same as the interaction (per unit length) 
between vortex lines in a type-II superconductor \cite{LP}.
In Fig.~\ref{fig:Eint}, this is compared with numerical results obtained on a lattice. 
For computation of 
the energy of a single soliton, on the other hand, the continuum approximation can only 
achieve the logarithmic accuracy (at small $\mu$), 
as there is a large contribution from $r \sim 1$. For reference, the 
energy of a single soliton with $\mu^2 = 0.1$, computed numerically, is $E_{sol} = 0.814$.

\begin{thebibliography}{99}
\bibitem{Dasgupta&Halperin} C. Dasgupta and B. I. Halperin,
  ``Phase Transition in a Lattice Model of Superconductivity,"
  Phys. Rev. Lett. {\bf 47}, 1556 (1981).
\bibitem{Skyrme} T. H. R. Skyrme,
  ``A non-linear field theory," Proc. R. Soc. A {\bf 260}, 127 (1961).
\bibitem{Skyrme2} T. H. R. Skyrme,
  A unified field theory of mesons and baryons, Nucl. Phys. {\bf 31}, 556 (1962).
\bibitem{soli} S. Khlebnikov, 
  ``Electron as soliton: Nonlinear theory of dielectric polarization," 
  arXiv:0710.0414.
\bibitem{Levin&Wen} M. A. Levin and X.-G. Wen.
  ``String-net condensation: A physical mechanism for topological phases,"
  Phys. Rev. B {\bf 71}, 045110 (2005) [arXiv:cond-mat/0404617].
\bibitem{Wen} X.-G. Wen,
  ``An introduction to quantum order, string-net condensation, and emergence 
  of light and fermions,"
  Ann. Phys. {\bf 316}, 1 (2005) [arXiv:cond-mat/0406441].
\bibitem{Wilson} K. G. Wilson, ``Confinement of quarks," Phys. Rev. D {\bf 10}, 2445 (1974).
\bibitem{Dirac} P. A. M. Dirac,
  ``Quantised  Singularities  in  the  Electromagnetic  Field,"  
  Proc. R. Soc. A {\bf 133}, 60 (1931).
\bibitem{Polyakov} A. M. Polyakov, 
  ``Compact gauge fields and the infrared catastrophe," Phys. Lett. {\bf 59B}, 82 (1975).
\bibitem{Bottcher&al} C. G. L. B\o{}ttcher, F. Nichele, M. Kjaergaard, H. J. Suominen, 
  J. Shabani, C. J. Palmstr\o{}m, and C. M. Marcus,
  ``Superconducting, insulating and anomalous metallic regimes in a gated two-dimensional 
  semiconductor-superconductor array,"
  Nature Phys. {\bf 14}, 1138 (2018) [arXiv:1711.01451].
\bibitem{Arnold} V. I. Arnold, {\em Geometrical Methods in the Theory of Ordinary
  Differential Equations}, 2nd Edition (Springer, 1988).
\bibitem{CRC} L. I. Berger,
  ``Properties of Semiconductors,'' in {\em CRC Handbook of Chemistry and Physics}, 
  106th Edition, John R. Rumble, ed. (CRC Press/Taylor \& Francis, Boca Raton, FL, 2025).
\bibitem{Mooij&Nazarov} J. E. Mooij and Yu. V. Nazarov,
  ``Superconducting nanowires as quantum phase-slip junctions," 
  Nature Phys. {\bf 2}, 169 (2006).
\bibitem{quantum_mech} S. Khlebnikov,
  ``Quantum mechanics of superconducting nanowires,"
  Phys. Rev. B {\bf 78}, 014512 (2008) [arXiv:0803.0975].
\bibitem{LP} E. M. Lifshitz and L. P. Pitaevskii, {\em Statistical Physics, Part 2}
  (Butterworth-Heinemann, 1980).
\bibitem{FR} D. Finkelstein and J. Rubinstein,
  ``Connection between Spin, Statistics, and Kinks,"
  J. Math. Phys. {\bf 9}, 1762 (1968).
\bibitem{Kogut&Susskind} J. Kogut and L. Susskind,
  ``Hamiltonian formulation of Wilson's lattice gauge theories," 
  Phys. Rev. D {\bf 11}, 395 (1975).
\bibitem{Jahani&Jacob} S. Jahani and Z. Jacob,
  ``All-dielectric metamaterials," 
  Nature Nanotech. {\bf 11}, 23 (2016).
\bibitem{Buttiker} M. B\"{u}ttiker,
  ``Zero-current persistent potential drop across small-capacitance Josephson junctions,"
  Phys. Rev. B {\bf 36}, 3548 (1987).
\bibitem{Bouchiat&al} V. Bouchiat, D. Vion, P. Joyez, D. Esteve, and M. H. Devoret,
  ``Quantum Coherence with a Single Cooper Pair,"
  Physica Scripta {\bf T76}, 165 (1998).
\bibitem{Mooij&al}
  J. E. Mooij, B. J. van Wees, L. J. Geerligs, M. Peters, R. Fazio, and G. Sch\"{o}n,
  ``Unbinding of Charge-Anticharge Pairs in Two-Dimensional Arrays of Small Tunnel Junctions,"
  Phys. Rev. Lett. {\bf 65}, 645 (1990).
\end{thebibliography}
\end{document}